\documentclass{endm}
\usepackage{endmmacro}
\usepackage{graphicx}
\usepackage{amsmath, amssymb, amscd, amsfonts}
\usepackage[mathscr]{eucal}
\usepackage{mathrsfs}
\usepackage{lscape}
\usepackage{subfigure}

\newcommand{\remove}[1]{}

\newcommand{\pa}[2]{\mathbf{P}_{#1}^{(#2)}}

\newcommand{\hp}[2]{\mathbf{H}_{#1}^{(#2)}}

\newcommand{\G}{\mathbf{G}}
\newcommand{\eg}{\textit{e.g.}}

\begin{document}

\begin{verbatim}\end{verbatim}\vspace{2.5cm}

\begin{frontmatter}

\title{Independent subsets of powers of paths, and Fibonacci cubes}

\author[Milan]{Pietro Codara\thanksref{myemail}\thanksref{supportedby}},
\author[Milan]{Ottavio M. D'Antona\thanksref{myemail}}
\address[Milan]{Dipartimento di Informatica, Universit\`{a} degli Studi di Milano,\\ I-20135, Milan, Italy}
\thanks[myemail]{Emails: \href{mailto:codara@di.unimi.it} {\texttt{\normalshape codara@di.unimi.it}},
\href{mailto:dantona@di.unimi.it} {\texttt{\normalshape dantona@di.unimi.it}}}
\thanks[supportedby]{Partially supported by ``Dote ricerca'' -- FSE, Regione Lombardia.}

\begin{abstract}
We provide a formula for the number of edges of
the Hasse diagram of the independent subsets of
the $h$\textsuperscript{th} power of a path ordered by
inclusion. For $h=1$ such a value is the number of edges of a Fibonacci
cube. We show that, in general, the number of edges of the diagram
is obtained by convolution of a Fibonacci-like sequence with itself.
\end{abstract}


\end{frontmatter}

\section{Introduction}\label{intro}
For a graph $\G$ we denote by $V(\G)$ the set of its vertices, and by $E(\G)$
the set of its edges.
\begin{defn}\label{def:h-path}\label{def:h-cycle}
For $n, h\geq0$, the \emph{$h$-power of a path}, denoted by $\pa{n}{h}$, is a graph
with $n$ vertices $v_{1}, v_{2}, \dots, v_{n}$ such that, for $1\leq i,j\leq n$, $i\neq j$,
$(v_i,v_j)\in E(\pa{n}{h})$ if and only if $|j-i|\leq h$.
\end{defn}
Thus, for instance, $\pa{n}{0}$ is the graph made of $n$ isolated nodes, and
$\pa{n}{1}$ is the path with $n$ vertices.

\begin{defn}\label{def:indipendent subset}
An \emph{independent subset of a graph $\G$} is a subset of $V(\G)$
not containing adjacent vertices.
\end{defn}

\noindent\textbf{Notation.} (i) We denote by $p_n^{(h)}$ the number of independent subsets of $\pa{n}{h}$.
(ii) We denote by $\hp{n}{h}$ the Hasse diagram of the poset of independent subsets of
$\pa{n}{h}$ ordered by inclusion, and by $H_n^{(h)}$ the number of edges of $\hp{n}{h}$.

%

\smallskip
In this work we evaluate $p_n^{(h)}$,
and $H_n^{(h)}$.
Our main result (Theorem \ref{th:main}) is that,
for $n,h\geq 0$, the sequence $H_n^{(h)}$ is obtained
by convolving the sequence $\underbrace{1,\dots,1}_{h},p_{0}^{(h)},p_{1}^{(h)},p_{2}^{(h)},\dots$
with itself.

Clearly, $\hp{n}{0}$ is the $n$-dimensional cube. Thus, on one hand,
our work generalizes the known formula $n2^{n-1}$ for the number of edges of the Boolean
lattice with $n$ atoms, obtained by the convolution of the sequence $\{2^n\}_{n\geq 0}$ with itself.
From a different perspective, this work could be seen as yet another generalization of
the notion of Fibonacci cube. Indeed, observe that every independent subset
$S$ of $\pa{n}{h}$ can be represented by a binary string $b_1 b_2 \cdots b_n$,
where, for $i=1,\dots,n$, $b_i = 1$ if and only if
$v_i \in S$. More specifically, each
independent subset of $\pa{n}{h}$ is associated with a binary string of length
$n$ such that the distance between any two $1$'s of the string is greater than $h$.
For $h=1$ the binary strings associated with independent subsets of $\pa{n}{h}$
are \emph{Fibonacci strings of order $n$}, and the Hasse diagram of the set of all such strings ordered
bitwise is a \emph{Fibonacci cube of order $n$} (see \cite{survey_fibo,fibocubes_enum}). Fibonacci cubes were
introduced as an interconnection scheme for multicomputers in \cite{hsu}, and their
combinatorial structure has been further investigated, \eg\, in \cite{klavzar,fibocubes_enum}.
Several generalizations of the notion of Fibonacci cubes has been proposed
(see, \eg, \cite{gen_fibo,survey_fibo}). As far as we now, our generalization, described
in terms of independent subsets of powers of paths ordered by inclusion, is a new one.

\section{The independent subsets of powers of paths}\label{sec:ind sub of paths}

We denote by $p_{n,k}^{(h)}$ the number of  independent
$k$-subsets of $\pa{n}{h}$.

\begin{lem}
\label{lem:p_nk formula}
For $n, h, k \geq 0$, $p_{n,k}^{(h)} = \binom{n-hk+h}{k}$.
\end{lem}
\begin{pf}
See \cite[Theorem 1]{hoggatt}, and \cite{arxiv}, where we
establish a bijection between independent $k$-subset of $\pa{n}{h}$ and $k$-subsets
of a set with $(n-hk+h)$ elements.\qed
\end{pf}
For $n,h\geq0$, the number of  all independent subsets of $\pa{n}{h}$ is

\medskip\centerline{
$p_n^{(h)}=\sum_{k=0}^{\lceil n/(h+1)\rceil}p_{n,k}^{(h)}=\sum_{k=0}^{\lceil n/(h+1)\rceil}\binom{n-hk+h}{k}\,.$
}

\begin{rem} Denote by $F_n$ the $n^{th}$ element of the Fibonacci sequence $F_1=1$, $F_2=1$, and $F_i=F_{i-1}+F_{i-2}$, for $i>2$. Then, $p_n^{(1)}=F_{n+2}$.
\end{rem}

\begin{lem}
\label{lem:p_n recurrence}
For $n, h \geq 0$,
$
p_n^{(h)} = \begin{cases}
n+1 & \text{if }\ n \leq h+1\,, \\
p_{n-1}^{(h)}+p_{n-h-1}^{(h)} & \text{if }\ n>h+1\,.
\end{cases}
$
\end{lem}
\begin{pf}
See the first part
of \cite[Proof of Theorem 1]{hoggatt}, or \cite{arxiv}.\qed
\end{pf}

\section{The poset of independent subsets of powers of paths}
\label{sec:hasse of paths}

\begin{figure}[h!]
  \centerline{\includegraphics[scale=0.6]{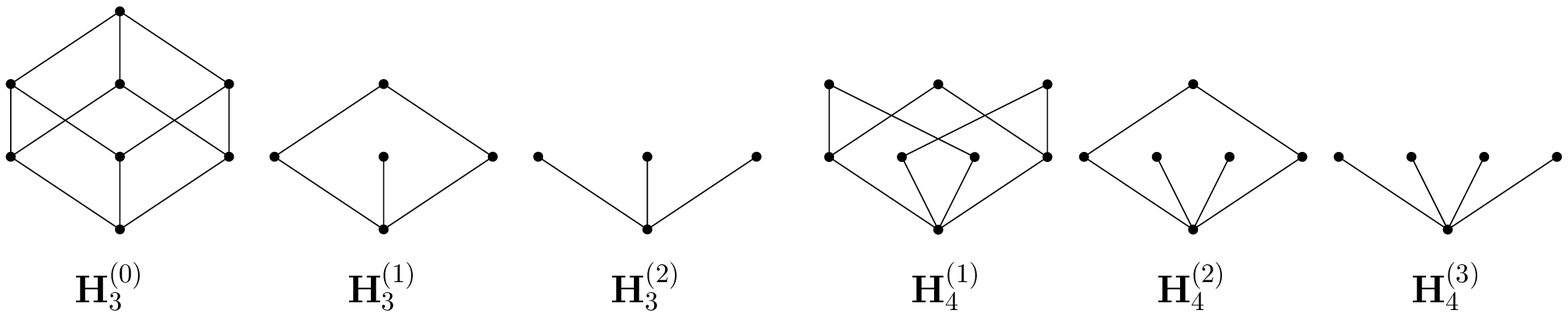}}
  \caption[Some $\hp{n}{h}$]
    {Some $\hp{n}{h}$.}
  \label{fig:P_3-4_h}
\end{figure}

Figure \ref{fig:P_3-4_h} shows a few Hasse diagrams $\hp{n}{h}$. Notice that,
as mentioned in the introduction, for each $n$, $\hp{n}{1}$ is a Fibonacci cube.

Since in $\hp{n}{h}$ each non-empty independent $k$-subset covers
exactly $k$ independent $(k-1)$-subsets, we can write

\begin{equation}\label{eq:H_nh formula}
H_n^{(h)}= \sum_{k=1}^{\lceil n/(h+1)\rceil}kp_{n,k}^{(h)}= \sum_{k=1}^{\lceil n/(h+1)\rceil}k{{n-hk+h}\choose k}\ .
\end{equation}

Let now $T_{k,i}^{(n,h)}$ be the number of independent $k$-subsets of $\pa{n}{h}$ containing the vertex $v_i$,
and let, for $h,k\geq 0$, $n \in \mathbb{Z}$,
$\bar{p}_{n,k}^{(h)} = \begin{cases}
p_{0,k}^{(h)} & \text{if }\ n < 0\,, \\
p_{n,k}^{(h)} & \text{if }\ n \geq 0\,.
\end{cases}$

\begin{lem}
\label{lem:T formula}
For $n,h,k \geq 0$, and $1\leq i \leq n$,
\[
T_{k,i}^{(n,h)}=\sum_{r=0}^{k-1}\, \bar{p}_{i-h-1,r}^{(h)} \ \bar{p}_{n-i-h,k-1-r}^{(h)}\,.
\]
\end{lem}
\begin{pf}
No independent subset of $\pa{n}{h}$ containing $v_i$ contains any of
the elements $v_{i-h}, \dots,v_{i-1},v_{i+1},\dots,v_{i+h}$.
Let $r$ and $s$ be non-negative integers whose sum is $k-1$. Each independent $k$-subset of $\pa{n}{h}$ containing $v_i$
can be obtained by adding $v_i$ to a $(k-1)$-subset $R\cup S$ such that

\noindent(a) $R\subseteq \{v_1,\dots,v_{i-h-1}\}$ is an independent $r$-subset of $\pa{n}{h}$;

\noindent(b) $S\subseteq \{v_{i+h+1},\dots,v_n\}$ is an independent $s$-subset of $\pa{n}{h}$.

\smallskip
Viceversa, one can obtain each of this pairs of subsets by removing $v_i$ from an independent
$k$-subset of $\pa{n}{h}$ containing $v_i$.
Thus, $T_{k,i}^{(n,h)}$ is obtained by counting independently
the subsets of type (a) and (b). Noting that the subsets of type (b) are
in bijection with the independent $s$-subsets of
$\pa{n-i-h}{h}$,
the lemma is proved.\qed
\end{pf}

In order to obtain our main result, we prepare a lemma.

\begin{lem}
\label{lem:somma tabelle T}
For positive $n$,
\[
H_n^{(h)} = \sum_{k=1}^{\lceil n/(h+1)\rceil}\sum_{i=1}^n T_{k,i}^{(n,h)}\,.
\]
\end{lem}
\begin{pf}
The inner sum
counts the number of $k$-subsets exactly $k$ times, one for each element of the subset.
That is,
$\sum_{i=1}^n T_{k,i}^{(n,h)} = k p_{n,k}^{(h)}$.
The lemma follows directly from Equation (\ref{eq:H_nh formula}).\qed
\end{pf}

Next we introduce a family
of Fibonacci-like sequences.

\begin{defn}\label{def:gen fibonacci}
For $h\geq 0$, and $n\geq 1$, the \emph{$h$-Fibonacci sequence}
$\mathcal{F}^{(h)}=\{F_n^{(h)}\}_{n\geq 1}$ is the sequence whose elements are
\[
F_n^{(h)} = \begin{cases}
1 & \text{if }\ n \leq h+1\,, \\
F_{n-1}^{(h)}+F_{n-h-1}^{(h)} & \text{if }\ n>h+1.
\end{cases}
\]
\end{defn}

\noindent From Lemma \ref{lem:p_n recurrence}, and setting
for $h\geq 0$, and $n \in \mathbb{Z}$,
$\bar{p}_{n}^{(h)} = \begin{cases}
p_{0}^{(h)} & \text{if }\ n < 0\,, \\
p_{n}^{(h)} & \text{if }\ n \geq 0\,,
\end{cases}$
we have that,
\begin{equation}\label{eq:F and p}
F_i^{(h)}=\bar{p}_{i-h-1}^{(h)}\,, \ \ \text{for each}\ \  i\geq 1\,.
\end{equation}

\noindent Thus, we can write
$\mathcal{F}^{(h)}=\underbrace{1,\dots,1}_{h},p_{0}^{(h)},p_{1}^{(h)},p_{2}^{(h)},\dots\ $.

\smallskip
In the following, we use the discrete convolution operation $\ast$, as follows.
\begin{equation}\label{eq:convolution}
\left(\mathcal{F}^{(h)}\ast \mathcal{F}^{(h)}\right)(n)\doteq \sum_{i=1}^n F_{i}^{(h)} F_{n-i+1}^{(h)}\,.
\end{equation}

\begin{thm}
\label{th:main}
For $n,h\geq 0$, the following holds.

\medskip\centerline{
$H_n^{(h)} = \left(\mathcal{F}^{(h)}\ast \mathcal{F}^{(h)}\right)(n)\,.$
}
\end{thm}
\begin{pf}
The sum $\sum_{k=1}^{\lceil n/(h+1)\rceil} T_{k,i}^{(n,h)}$
counts the number of independent subsets of $\pa{n}{k}$ containing $v_i$.
We can also obtain such a value by counting
the independent subsets of both $\{v_1, \dots, v_{i-h-1}\}$, and $\{v_{i+h+1}, \dots, v_{n}\}$.
Thus,  we have:

\smallskip\centerline{
$\sum_{k=1}^{\lceil n/(h+1)\rceil} T_{k,i}^{(n,h)} = \bar{p}_{i-h-1}^{(h)}\, \bar{p}_{n-h-i}^{(h)}\,.$
}

\medskip\noindent Using Lemma \ref{lem:somma tabelle T} we can write

\smallskip\centerline{
$H_n^{(h)} = \sum_{k=1}^{\lceil n/(h+1)\rceil}\sum_{i=1}^n T_{k,i}^{(n,h)} =
\sum_{i=1}^n\sum_{k=1}^{\lceil n/(h+1)\rceil} T_{k,i}^{(n,h)} =
\sum_{i=1}^n \bar{p}_{i-h-1}^{(h)}\, \bar{p}_{n-h-i}^{(h)}.$
}

\medskip\noindent By Equation (\ref{eq:F and p}) we have
$\sum_{i=1}^n \bar{p}_{i-h-1}^{(h)}\, \bar{p}_{n-h-i}^{(h)}=\sum_{i=1}^n F_{i}^{(h)} F_{n-i+1}^{(h)}\,.$
By (\ref{eq:convolution}), the theorem is proved.\qed
\end{pf}

Further properties of coefficients $H_n^{(h)}$, and $p_n^{(h)}$ are discussed in \cite{arxiv}.
Moreover, in \cite{arxiv} we investigate the case of powers of cycles, and its connection with Lucas cubes.

\end{document}